RESEARCH ARTICLE                                                                 OPEN ACCESS

# Achievement of Objectives of Library Information Management: Result of Right Structuring of Library Network System

## Vikash Prajapat [1], Rupali Dilip Taru [2]

[1] Ph.D Scholar, Department of Library and Information Science, Nirwan University-Jaipur - India,
[2] Assistant Professor, Department of Management Studies, Bharati Vidyapeeth (Deemed to be University)
Department of Management Studies (Off Campus),
Navi Mumbai, Maharashtra - India

**ABSTRACT**
The world is transforming through a revolution and development in the progression of information and its broadcasting. The number of research journals, books and reports being published the world over has been increasing phenomenally. Currently, about five lakh books, one lakh periodicals, lakhs of patents, thousands of standards and numerous other types of documents are being published every year. Moreover three crore subject themes have been introduced and published all over the globe. And it's practically difficult for any individual library system to purchase and have information about each single book, journals and meet their various demands of the end user community. No library can possibly provide efficient services with its age old manual operations. The use of computers in libraries has been of great help in the acquisition, serial control, cataloguing, circulation besides keeping tracks of stocks and users. With the increasing use of digital media and computer devices with other communication facilities, broadcasting of information and sharing of resources is becoming more and more significant and thus the inter-connectivity with and on the other libraries network to fulfill the ever increasing end users demands of wide range of informational resources. [13]And this paper explains how right structuring of Library Networks support to achieve the objectives of Library Information Management. For analyzing the feedback received, we created a dataset is a range of contiguous cells on an Excel worksheet containing data to analyze its mean and standard deviation. In addition to computing the alpha coefficient of reliability, where Cronbach's alpha is a measure of internal consistency. The hypothesis was tested using SPSS to obtain co-variances by going to Analyze – Correlate – Bivariate, A Likert-type scale was prepared and used to capture the feedback using the feedback survey Total 90 library staff were asked to complete this survey through questionnaires. We also checked the dimensionality of the scale and factor analysis to cross check direction of relationship between all four variables.

## I. INTRODUCTION

In India, the functioning of standardization of and modernization of libraries and formation of library networks reached and attended quite late in implementation of its establishment. NISSAT made it possible with efforts to establish CALIBNET in 1986; DELNET in 1988 and various different like ADINET, PUNENET, BONET subsequently. The UGC formed INFLIBNET in 1988. INSDOC helped the establishment MALIBNET in 1993. Further to these networks structures and various library networks entered into reality and now support a centralized database of library information management, and available to be retrieved and accessed by its handler libraries for the aim of resource sharing various information thought the store database. Implementation of modern infrastructural information technology and relative application integration has reached and brought out the histrionic changes in the library and information field. With information technological advancement libraries and information centres around the world have digitalized their library procedures of day to day activities, formed and developed relational databases for collective use on computer information networks.[8] In addition to improving services and functional operations for efficient performance libraries have also been force to use of advance and effective structural technological digital networks with an objective to maximize utilization of collected resources and facilities. The library and information networks and its strategically structuring operations have potential to enhance library and information services in various possible ways. It low down the price of information resource services in the structural network environment with the distributed mode. Such structured based information resources enables libraries to availed required base facilities to the end users excluding the drawbacks of exploration as per size, distance and language barriers within these entities among them.[9] With progression in library networks based on technological advancement, the prominence has encouraged from the set-ups of networks as physical units to the collective resources accessible through the structured integrated need based networks and its system of accessibility and distribution of library and information network and its management.[1] Such technological developed network provider and accessible resources include catalogues of library properties of holding database, research journal, scholarly articles, graphical and electronic text, invented images detailing, video and audio folders and files, informational scientific and modern technical database, etc. This phase commences library and information networks, their aim and fundamental features. It delivers classification of library and information networks and recognizes online shared databases, computer hardware structure, integrated software infrastructure, database accessible networks and human interaction network and agreed members of a library network these are the important components. On such phase collaboration of these different components describes tangible and financial limitation, information literature explosion and improved awareness and mandate of demand from the library users as main requirement of libraries that directed to the evolution and expansion of library networks in the world. It offers an ephemeral past and development of library networks. A library and information network can propose a various number of facilities and services subject to its aim and objectives, kind of demand from the united or collaborative libraries. The paper explains on the relationship and signified role of right structuring of library network system in achieving the objectives of library information management. Where a library can deliver the require services to its members with comprehensive account of the library & information networks and their facilities. The operative access to collective possessions of library resource database through digitalized databases of several member associations;

Objectives of Library and Information Networks: The main reason for such structured network layout and its integration to share the data and information service facilities to library users and members through utilization of such resources and participation on various platforms. Structuring of right library network system improve resource utilization and service level at the individual libraries by providing automation facilities in the listed areas like cataloguing of books, non-book materials and catalogue production, serials, Acquisition and funding resource accounting, Circulation, member subscriptions, implementing electronic services in the libraries for fast communication of information resources and to promote the resource sharing and supportive coordinated activities among the libraries on conditions that well-organized and consistent revenues of resource allotment, stimulating, authorizing and organizing research and training programme for library staff and network members, Entrance to national and international collections, communication association concluded periodical and inter-personal messages of library network which is the standards and uniform with guidelines in techniques, methods, procedures of communication and integration of hardware and software combination to provide integrated services. [2] So in order to provide the integrated coordinative support on regional, national and international networks, library right networking structure plays an vital role in exchange of information and documents for the use of library staff and member subscribers. [7]

## II. STRUCTURE OF LIBRARY INFORMATION NETWORKING:

TYPES OF NETWORKS: There are three main types of computer networks[10]:

**Local Area Network (LAN):** LAN means connected computer systems and electronic devices that share information resources over a transmission and communication media. LAN connection can use a single printer and a file server. The LAN network can be within a same building or a same campus network.

**Metropolitan Area Network (MAN):** Various attempts were being completed to bring out this type of network system in metropolitan areas such Bangalore, Madras, Delhi, Calcutta etc.

**Wide Area Network (WAN):** this is a large-scale network, connecting offices in various cities and countries are known as WAN, specifically structured to interconnect data broadcast devices over wide geographical areas.

## III. CATEGORIES OF NETWORK LIBRARY

The structure of a network will depend upon the purpose for which libraries use it. A number of libraries should join the network that offers the facilities to make their functioning simpler, better and more cost effective (i) General network, (ii) Specialized network, (iii) Metropolitan network, (iv) Countrywide network. Based above stated functions of a library network different models of networks can be formed and implement:

**Star Network:** A star network where, one of the associates maintains operates and managed a centralized integrated bibliographical resource database. Such database involves all the bibliographical records of each and every member libraries. In this configuration node P maintain

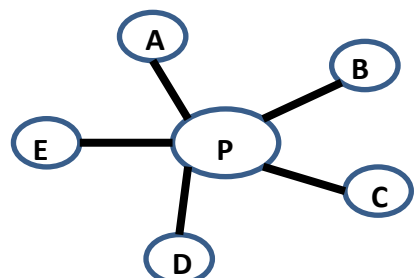

**Image 1: Star Network**

Manage and update the bibliographical databases and all other members A, B, C, D, and E utilizing these resources.

**Hierarchical Network:** in this kind of network each member shares resources nearby; participating library at the lower level forward the unsatisfied requirements to the higher level participating library. Obliged to check other centres to locate required materials at `library of last resort' to referred unsatisfied requirements at lower level

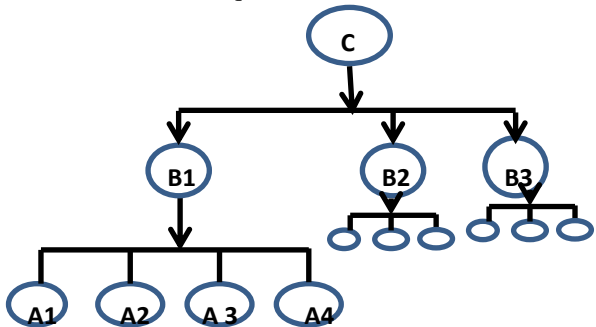

**Image 2:  Hierarchical Network**

In this configuration network, members $A_1$, $A_2$, $A_3$, $A_4$ are mostly satisfied through the resources allocation from respective libraries. Unsatisfied queries passed on to the higher level resource centre B, and finally in remaining unsatisfied requests are referred to the library of last resort C, which may be obliged to patterned other hubs $B_2$ and $B_3$ to locate essential ingredients.

**Distributed Network:** In distributed network where all network members hold in principle of overall complete decentralized pattern. Various different resources may share with one another. In this

configuration all library users A, B, C, D, E, have different resources requisition

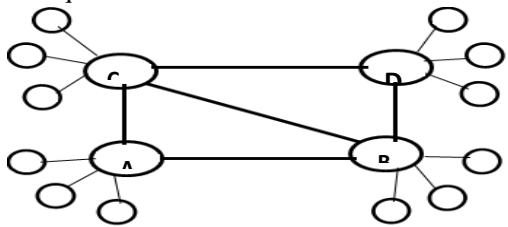

**Image 3:  Distributed Network**

Where A, B, C and D are themselves a network, it could be the grouping of star category type of network. This is adapted model of a distributed network as distribution among one another.

## IV.     NETWORK                 PROTOCOLS     STANDARDS

On network all computers are communicating to each other, and protocol plays a very important role which is actually the set of rules by which computers can communicate. The protocols are set of rules which govern the flow of data transmission from one computer to another. Due to this set of instruction different types of computer are able to communicate effectively based on agreed standards define in protocols. [3]

**(A) Network Standards**: International agreement sets these standards or specific manufacturer become the dominant in this networking filed. The most three important authorities take decisions in setting such network standards in sector of computer networking. [6]

- CCITT: Comite Consultatif International Telegraphique et Telephonique
- ISO: Open Systems Interconnection
- IEEE: Institute of Electrical and Electronics Engineers

**CCITT: (Consultative Committee for International Telephony and Telegraphy)** are accountable for the V, X and I standards.

**ISO: (International Standard Organization)** developed a seven layer model which known as Open Systems Interconnection i.e. OSI model this describe the direction in which computer networks operational process work among the network devises. This model operates through physical connections at layer one till end at applications at layer seven in between this there are five more layers named data link layer, network layer, transportation layer session layer and presentation layer. The goal of this model is to support grow `open' standards.

**IEEE: Institute of Electrical and Electronics Engineers** are an American authority who advanced, developed and recommended such standards for the physical connection of computers in network.[5]

**(B) Protocols controlling data transfer over the network:** In the initial year of 1960s when researchers tried to link workstations with each other they assumed of several methods to administer the movement of data files from one workstation to another in network. Here is the network protocol:

**(i) NCP (Netware Core Protocol):** Developed by Novell Netware, it succeeds the movement of data amongst Netware Clients and File Server for determined efficacy.

**(ii) TCP/IP (Transmission Control Protocol/Internet Protocol):** mechanisms based on the standard of packet switching which includes transgression the records into lesser amounts and directing these record portions endways of the network. Each packet is involved with facts of source and destination of that data packet; in this result many packets travailing and flowing through the same network system in link with every data packets successively reach at its correct destination.

**(iii) IPX (Internet work Packet Exchange):** It managed data authentication and addressing accountabilities for movement of data between various and diverse networks for example between Netware Networks model and other standard networks. This adds a header to the information acknowledged from TCP Protocol holding the address of the endpoint destination and then such headers passes the data to communicate with the Network Interface Card for transmission of data.

**(iv) SPX (Sequenced Packet Exchange):** It confirms and checks the correctness of the information data by accessing by NOS (Network Operating System). It also confirms the accurateness of the remote workstation to where the data communication performing its initial communications with the destination workstation.

**(C) Standards & Protocols of Library and Information Science Field:** with the appropriate tools and techniques resource sharing technology cannot work effectively and can developed fast. Such techniques involve the establishment of standard bibliographic databases though, the acceptance of changed bibliographic standards generates inconsistency difficulties and that acts by means of a main obstacle in the use of bibliographic and linked info. The decided standards, set rules of protocols and the essential software that convert related data from one format to another with maintaining the international standard.

**(i) Bibliographic Standards:** Layout compatibilities are essential for electronic index documents and these are actuality uniform by the International Standards Organization. Conservation of bibliographic standards consistently is important for rapid discussion of registers, Standards have been developed by the International Standards Organization and National Information Standards Organization of the US besides a number of national standards association for constructing customary standard bibliographic and information products. In this protocol provides a mechanism, applying client's server model for transmitting and controlling queries and effect collections. The protocol is expected to be POST (Open System Interconnection) founded and is competent to interconnect with all sorts of electronic hardware, application software and transmission services used. OSI delivers inter-connectivity and inter-operability in direction to interconnect with all sorts of hardware and agree transmission and declaration of documents between them. The protocol has been deliberate to reassure software developers and associates of the library management system [12] and the info communal to approve the protocol and integrate it into library info link systems.

## V. OBJECTIVES
To study the impact of right structuring of library network system has direct relationship to achieve the objectives of Libraries and Information Management.

## VI. HYPOTHESIS

H 0: r = 0 there is no relationship between 'having right structuring of library network system and to achieve the objectives of Libraries and Information Management'.

H 1: r ≠ 0 there is a significant relationship between 'having right structuring of library network system and to achieve the objectives of Libraries and Information Management'.

## VII. METHODOLOGY
The subsequent primary parameters were estimated – Communication, Timelines, Quality and Overall Feedback to 90 library administrative staff were analysed this feedback survey. The surveys responses were compiled together. Where the significance level p value= 0.05. The overall survey respondents will be the sample for the study. The total population includes all the employees of library staff of different university which availed from conferences events. A Likert type scale prepared to capture the response using the feedback extending from (1) Excellent, (2) Good (3) Neutral (4) Bad (5) Worst with these concluded closed ended questionnaire. This study was collectively shared between the library staff with the support of online survey tool link. This tool was used to record and all the feedback replies. All the answers received from the past six months. Same has been analysed for the purpose of this research study with the support of statistical methods. A total of 90 responses were collected and collaborated in excel. For analysing the feedback received, created a dataset is a range of contiguous cells on an Excel worksheet containing data to analyse its mean and standard deviation to calculate the mean of a dataset in Excel, function, where Range is the range of values has been summarized as below:

| | Variables | Count | Mean | St Dev |
|---|---|---|---|---|
| 1 | Information services to users | 90 | 4.626 | 0.551 |
| 2 | Technical services to member libraries | 90 | 4.703 | 0.587 |
| 3 | Management service to the network administration | 90 | 3.912 | 0.877 |
| 4 | Categorical based services to network groups | 90 | 3.978 | 0.989 |

**Table 1: Analysis -mean and standard deviation**

The hypothesis was tested using SPSS to obtain co-variances by going to Analyse–Correlate–Bivariate; we also checked the dimensionality of the scale and factor analysis on collected data from respondents.

| | | q1 | q2 | q3 | q4 |
|---|---|---|---|---|---|

| q1 | Covariance | **1.17** | 0.55 | 0.56 | 0.64 |
| q2 | Covariance | 0.55 | **1.01** | 0.59 | 0.71 |
| q3 | Covariance | 0.56 | 0.59 | **1.17** | 0.71 |
| q4 | Covariance | 0.64 | 0.71 | 0.71 | **1.28** |

Table 2: Computing Covariance

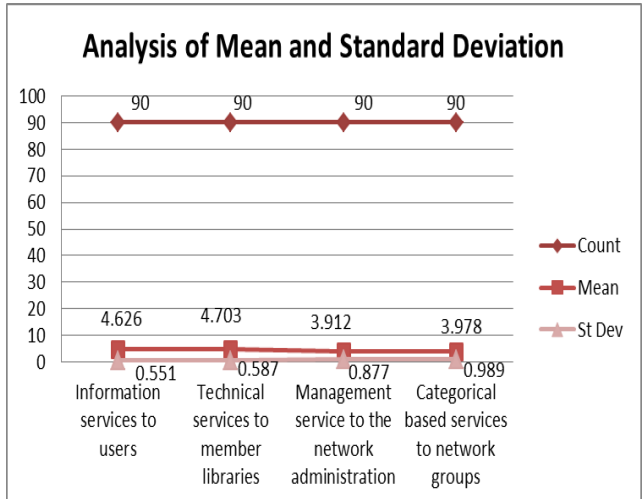

Graph 1: Analysis -mean and standard deviation

Correlation is a bivariate analysis that procedures the strength of suggestion between two variables and the direction of the relationship. In regards of the influence of association, the significance of the correlation coefficient differs between '+1' and '-1'. In addition to computing the alpha coefficient of reliability to analyse the data generated from 90 completed questionnaires. A "high" value for alpha does not imply that the measure is one-dimensional technically; Cronbach's alpha is not a statistical test. It is a coefficient of reliability (or consistency) that is, how closely linked a set of items is as a group. If, in addition to measuring internal consistency, just to provide evidence that the scale in question is one-dimensional, additional analyses performed here. Exploratory type of factor analysis is one practice to verifying the dimensionality status. In SPSS, we obtain covariance's by going to Analyse – Correlate – Bivariate. And provided input of q1, q2, q3 and q4 to the variables box and click on options, under statistics, check Cross-product deviations and covariance. Where following a summarized version of the output as Table no. 2 where the diagonals (in bold) are the variances and the off -diagonals are the covariance's. We only consider the covariance's on the lower left triangle because this is a symmetric matrix.

Cronbach's alpha can be printed as a purpose of the number of examination substances and the normal inter-correlation between the substances. Below, for theoretical resolutions, we show the formulation for the Cronbach's alpha:

$$\alpha = \frac{N\bar{c}}{\bar{v} + (N-1)\bar{c}}$$

Formula 1: Cronbach's alpha

Here is identical to the quantity of items, is the average inter-item covariance between the items and equivalents the average inconsistency. We can see from this formula that if we increase the number of items, it increases Cronbach's alpha. And with this if the average inter-item correlation is low then the alpha value will be low. And as the average inter-item correlation increases the increase in Cronbach's alpha will found, in calculations the number of items constant, where n=4 is equal to the number of objects ‾c is the ordinary inter-item covariance between the objects and ‾v matches the average alteration. Additional to these particulars as the information from the stated table above, we can analysed each of these mechanisms via the following calculations:

$$\bar{v} = (1.166+1.011+1.167+1.281) / 4 = 4.625 / 4 = 1.156$$

$$\bar{c} = (0.547+0.564+0.590+0.637+0.710+0.714) / 6$$

$$= 3.762/ 6 = 0.627$$

$$\alpha = \frac{4\ (0.627)}{(1.156)+ (4-1)\ (0.627)}$$

$$\alpha = 2.508 / 3.037$$

$$\boxed{\alpha = 0.826}$$

Equation 1: Value of Cronbach's alpha

The outcomes match with values of below explained SPSS obtained Cronbach's Alpha of 0.826, to check the dimensionality of the scale using factor analysis. For this example, we will use a dataset that contains our test items – q1, q2, q3and q4. To compute Cronbach's alpha for all four items – q1, q2, q3, q4 –use the reliability command in SPSS:

RELIABILITY
/VARIABLES=q1 q2 q3 q4.

| | N | % |
|---|---|---|
| Valid | 90 | 100 |
| Excluded* | 0 | 0 |
| Total | 90 | 100 |

Table 3: All Variables, Processing Summary

*List wise deletion based on all variables in procedure

Reliability Statistics

| Cronbach's Alpha | N of Items |
|---|---|

| 0.826 | 4 |
|-------|---|

**Table 4: Reliability Statistics**

The alpha coefficient for the four items is **0.826**, suggesting that the items have relatively high internal consistency.

CHECKING DIMENSIONALITY
With computation of alpha coefficient of reliability additionally we investigate the dimensionality of the scale which might require in this study. For that we use the **factor** command:

FACTOR ANALYSIS

FACTOR
/VARIABLES q1 q2 q3 q4

## Communalities

|     | Initial | Extraction |
|-----|---------|------------|
| q1  | 1.000   | .585       |
| q2  | 1.000   | .721       |
| q3  | 1.000   | .685       |
| q4  | 1.000   | .714       |

**Table 5: Communalities**

Extraction Method Principal Component Analysis
Total Variance Explained

| Component | Initial Eigenvalues | | | Extraction sums of Squared Loadings | | |
|-----------|-------|------------|-------------|-------|------------|-------------|
|           | Total | % of Variance | Cumulative % | Total | % of Variance | Cumulative % |
| 1 | 2.706 | 67.654 | 67.654 | 2.706 | 67.654 | 67.654 |
| 2 | .541 | 13.531 | 81.185 | | | |
| 3 | .400 | 10.006 | 91.191 | | | |
| 4 | .352 | 8.809 | 100.000 | | | |

**Table 6: Total Variance Explained**

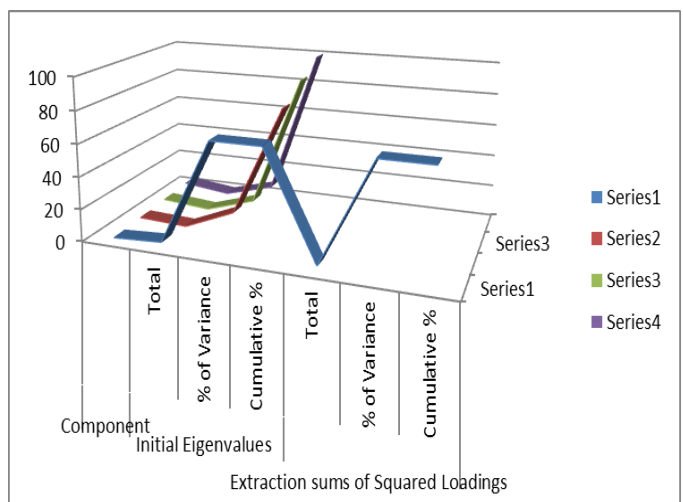

**Graph 2: Total Variance & Communalities%**

## Component Matrix*

|     | Component |
|-----|-----------|
|     | 1 |
| q2  | .849 |
| q4  | .845 |
| q3  | .828 |
| q1  | .765 |

**Table 7: Component Matrix**

Method of Extraction: a.1 Component Extracted, Principal Component analysis

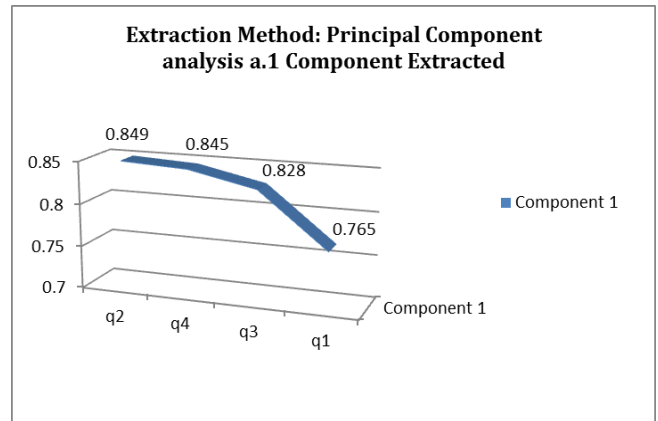

**Graph 3: Component Matrix***

## VIII. RESULTS AND DISCUSSION

Thought SPSS, we obtain covariances by analysing – Correlate –Bivariate. And provided input of q1, q2 , q3 and q4 which are based on variables i.e. are direct important functions of library information management based on library network system, with the Cross-product deviations and covariance's condensed version of the output where the diagonals in bold are the variances and the off -diagonals are the addition of (1.166: 1.011: 1.167: 1.281) covariance's. And considered the covariance's addition of (0.547: 0.564: 0.590: 0.637: 0.710: 0.714) on the lower left triangle because this is a symmetric matrix. Where covariance shows us the direction of the relationship between all four variables is very less. And all variable are equal values in their functionalities. The alpha coefficient for the four items is 0.826, suggesting that the Four items have relatively high internal consistency. Where a reliability coefficient of .70 or higher is considered "acceptable" in this study. Looking at the table labelled Total Variance that the eigenvalue for the first factor is quite a bit larger than the eigenvalue for the next factor 2.7 versus 0.54. Hence, the first factor accounts for 67% of the total variance. This recommends that the measure objects are unidimensional in nture. And the p-value measures the probability of getting a more extreme value than the one we got from the experiment. And here we have significance level is 0.05 p-value, which is less than alpha 0.826, so we reject the null hypothesis and analysis and testing proven that the right structuring of library networks system variables has direct impact to achieve the Objectives of Library Information Management. And accept the alternate hypothesis which noted that the essential functions should include the promotion of resource sharing, creation of resource sharing tools like union catalogues, explanation of achievements and adoption of international standards for creation of registers consistently and transfer of forms. Keeping these things in view, functions of a typical library network might fall into the subsequent three categories which are variables of formed questionnaire of this study, where (i) Information services to users, (ii) Technical services to member libraries, (iii) Management service to the network administration, (iv) Categorical based services, overall Information and technical services are goal-oriented, i.e. to complete the main goal of the network system. Functions that attend the handlers directly, i.e. information services to users are: **Information services**. Inter-library service which means, that each member library may require having the additional facility to photocopy for delivering the documents related resources. **Reference and Referral service where** member library may necessitate the competence of having devoted enquiry calls or E-mail services. **Access to databases:** to formulate reviewing bibliography for the purpose of glancing to know whether or not a manuscript is accessible. In **Technical Services:** Functions that serve the libraries i.e. the technical services are: Co-operative assortment improvement programme, Procedural processing involved in attainment, Classification and other resources of proceeds to recognize and to localize documents, Rotation regulator system. [4] **Management services** the management services, i.e. purposes that support the network administration are to establish an operational classification that implement the purposes stated above, Evaluation of the networks. This is done through collection of statistics, analysis of performance of the network, user studies. **Training activities**. [11] These are done through staff development programmes, user oriented programmes, Cost analysis it involves determination of costs, fees to be collected, etc. distribution of budgets. The library networks can be grouped under the following categories based on their services and activities like Umbrella and Supermarket, Bibliographic Utility Networks, Online Search Service Networks, Service Centre Networks for Subscription to Electronic Resources. Five major services of a library and information network are as follows Cooperative Cataloguing, Database Services, Document Delivery Services, Inter Library Credit, Collective Acquisition of Resources, and Consortium Procurement. Some of the important features of software are User friendly, Provision to on-line interaction, while editing recording, recovering and in data management, Provision to obtain outputs in various formats, Compatibility and portability. The resources so that unnecessary wastage of limited finance with them can be circumvented through right structured library network system. However, not all networks follow to the essential functions of library networks.

## IX. LIMITATIONS
In network development structure a network system may fail in the initial levels if there is inadequate planning or insufficient funds are not available.

## X CONCLUSIONS
A large count of library centres and information centre units are developing networks from the recent periods. The advent of computer networking as an accepted part of the library structure and infrastructure has had a very significant impact on the way in which library and information systems are perceived. India is thus on the beginning to a innovative period of computer communication networks both for common determinations and for library and information drives. The delegation of information and accessibility of database resources must be defined and structured. Moreover, a common memorandum of agreement signed by the participating libraries at the institutional level is essential for the success of a network venture and its transparency on administration of library information management system. On a more applied level, kind of catalog data must be in a standard, system of machine readable form for it to be collective and exchanged. Networking structure must define the access policy control, backup facility, online payment structure and error handling procedure, finally, a continuous flow of external assistance is crucial for the network's survival and internal management to keep operational and functional structure well in order.

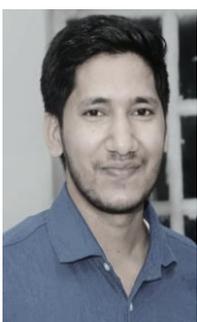

**First Author. Prajapat Vikash Jr.** (M' 98489397), this author became a Member (M) of IEEE in 2022, received the B.Sc. degree in mathematics from Maharaja Ganga Singh University, Bikaner, Rajasthan in 2018. He holds B. Lib and M. Lib degree in Library and Information Science from Shri JJT University, Jhunjhunu, Rajasthan in 2020 and 2021 respectively. He is currently pursuing the Ph.D. degree in Library and Information Science at Nirwan University, Jaipur Rajasthan, India.
From 2018 to till 2022, he is a Library Assistant With the Institute of Library and Information Science, Shri JJT University, Jhunjhunu, Rajasthan. His research interest includes the development of library information management system and identification of Library Network System, fundamental study of Library Information Science and Consortia of Digital Libraries and Information Networks.

Mr. Prajapat have participated in the 2nd Edition of the International Conference on Knowledge Management in Higher Education Institutions (ICKHI 2022), Organized by Central Library, Manipal University Jaipur in association with Library & Learning Resources Center, University of Dubai, held online on 11th & 12th April 2022 and presented the paper titled "Consortia of Digital Libraries and Information Networks: National and International"

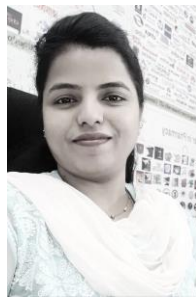

**Second Author. Dr. Taru Rupali** (M' 98489308). (Birthplace Pune, India.10/08/1981). This author became a Member (M) of IEEE in 2022, received the M.Com, MBA and LLM degrees in Master of Commerce and Management, Master of Business Administration and Master of Law from Mumbai University, Punjab Technical University and Shri JJT University in 2005, 2014 and 2021 respectively. She holds professional degree of PGDHRM and PGCSIR from Welingkar Institute of Management Mumbai and Asian School of Cyber Law, Pune respectively. In the year 2017, She has awarded with Philosophy of Doctorate in Management from, Shri Jagdishprasad Jhabarmal Tibrewala University, Jhunjhunu Rajasthan, India. She is now with Bharati Vidyapeeth (Deemed to be University) Department of Management Studies (Off Campus) Assistant Professor. And associated as Ph.D. Research Co-Guide in Nirwan University, Jaipur, Rajasthan. She is the author of two books 'SPSS for analytics' and 'DSS based on Info-Tech Structure of MIS' in year 2022. Research interest area is Management Information System and International and Comparative Law. Dr. Taru is sub-editor for Entire Research (ISSN 0975-5020) Global Human Research and Welfare Society. And awarded for Remarkable Contribution in field of Education in the 2nd International Conference in Dubai, Organized by Shri J.J.T. University and International Council of Jurists, London (UK).